\documentstyle[12pt]{article}

\newcommand{\be}{\begin{equation}}
\newcommand{\ee}{\end{equation}}
\newcommand{\bea}{\begin{array}}
\newcommand{\ea}{\end{array}}
\newcommand{\beqa}{\begin{eqnarray}}
\newcommand{\eeqa}{\end{eqnarray}}
\newcommand{\bean}{\begin{eqnarray*}}
\newcommand{\eean}{\end{eqnarray*}}
\newcommand{\eqn}[1]{(\ref{#1})}

\newcommand{\nn}{\nonumber}

\def\up#1{\leavevmode \raise.16ex\hbox{#1}}

\def\sqr#1#2{{\vcenter{\vbox{\hrule height.#2pt
	\hbox{\vrule width.#2pt height#1pt \kern#1pt
	  \vrule width.#2pt}
	\hrule height.#2pt}}}}

\newcommand{\jou}[4]{{\rm #1 }{\bf #2}, #3 \up(19#4\up)}

\newcommand{\gapproxeq}{\lower .7ex\hbox{$\;\stackrel{\textstyle >}{\sim}\;$}}
\newcommand{\lapproxeq}{\lower .7ex\hbox{$\;\stackrel{\textstyle <}{\sim}\;$}}

\def\thebibliography#1{{\bf REFERENCES\markboth
 {REFERENCES}{REFERENCES}}\list
 {[\arabic{enumi}]}{\settowidth\labelwidth{[#1]}\leftmargin\labelwidth
 \advance\leftmargin\labelsep
 \usecounter{enumi}}
 \def\newblock{\hskip .11em plus .33em minus -.07em}
 \sloppy
 \sfcode`\.=1000\relax}

\begin{document}

\title{\hfill $\mbox{\small{
$\stackrel{\rm\textstyle DSF-T-97/25\quad}
{\rm\textstyle hep-ph/9706300\quad\quad}$}}$ \\[1truecm]
Fermion Mass Matrices in terms of the Cabibbo--Kobayashi--Maskawa Matrix
and Mass Eigenvalues} 
\author{D. Falcone$^{*\dag}$, O. Pisanti$^*$, and L. Rosa$^{*\dag}$}
\date{$~$}

\maketitle

\thispagestyle{empty}

\begin{center}
\begin{tabular}{l}
$^*$  Dipartimento di Scienze Fisiche, Universit\`a di Napoli,\\
~~Mostra d'Oltremare, Pad.19, I-80125, Napoli, Italy; \\
$^{\dag}$ INFN, Sezione di Napoli, Napoli, Italy.\\ \\
\small\tt  e-mail: falcone@na.infn.it  \\
\small\tt e-mail: pisanti@na.infn.it  \\
\small\tt e-mail: rosa@na.infn.it 
\end{tabular}
\end{center}

\begin{abstract}
A parameter free, model independent analysis of quark mass matrices is
carried out. We find a representation in terms of a diagonal mass matrix
for the down (up) quarks and a suitable matrix for the up (down) quarks,
such that the mass parameters only depend on the six quark masses and the
three angles and phase appearing in the Cabibbo-Kobayashi-Maskawa matrix.
The results found may also be applied to the Dirac mass matrices of the
leptons. 
\end{abstract}

\bigskip


\newpage

Many attempts have been made to connect the quark mixing matrix with the
quark mass matrices introducing extra symmetries (or Ans$\ddot a$tze) to
cast the mass matrices in some particular form \cite{frit}. Branco, Lavoura
and Mota \cite{brla} have been able to show that for three families the
Nearest-Neighbor Interactions (NNI) form of mass matrices corresponds to a
choice of basis. Indeed, within the {\it Standard Model} \cite{stmo}, the
NNI form can be obtained by applying a particular transformation to the
fermionic fields without observable consequences. Relying on their result,
in ref. \cite{haro,koid,taka} the problem of finding mass matrices for the
fermions, as a function of physical parameters only, has been addressed.
Due to the NNI form they get very complicated relationships between mass
matrices parameters and the Cabibbo-Kobayashi-Maskawa matrix, $K$
\cite{cabi}. In particular, Harayama and Okamura \cite{haro} obtained
formulae in which two arbitrary phases (that is to say not determined by
physical parameters) still remain. Koide \cite{koid} showed that the two
phases can be eliminated by a change of phases of matrix elements. Finally
Takasugi \cite{taka} investigated the connections between NNI basis and the
USY (Universal Strength for Yukawa couplings) form of Yukawa coupling
\cite{brsi}, leaving for future works the problem of expressing quark mass
matrices in terms of physical parameters. 
 
In this paper we concentrate on this last problem. Using a particular basis
for quark (lepton) fields we find a representation of mass matrices in
which there are exactly ten free parameters, nine moduli and one phase. In
this basis it is possible to obtain relatively easy expressions for the
Cabibbo-Kobayashi-Maskawa matrix elements and, more interestingly, it is
possible to invert these relations linking the mass matrices with the
physical parameters. We quote exact and approximate formulae. We analyze
the quark-phase conventions and determine the expression for the observable
phase appearing in the mass matrices. We conclude with brief final remarks.

In what follows we concentrate on the mass and the weak-charged-current
terms of the {\it Standard Model} Lagrangian \cite{stmo}. We write them as
follows: 
\be
L=\bar{u}^0_L\tilde{M}_{u}u^0_R+\bar{d}^0_L\tilde{M}_{d}d^0_R+
g\bar{u}^0_L\not\!\!{W^+}d^0_L+h.c. 
\ee
(summation over family indices is intended).

It is possible to perform, with no physical consequences \cite{brla}, the
following transformations on the quark fields (a similar argument applies
to leptons): 
\be
\left\{ \bea{ c }  u^0_L=U u'_L \\  u^0_R=V_u u'_R  \ea \right. ~~~
\left\{ \bea{ c }  d^0_L=U d_L \\  d^0_R=V_d d_R  \ea \right.,
\label{bas1}
\ee
where the only constraint on the matrices $U,V_u,V_d$ is that they must be
unitary. We choose $U$, $V_d$ and $V_u$ so to have 
\be
L=\bar{u}'_L \hat{M}_{u} {u'}_R+\bar{d}_L{M}_{d}d_R+g\bar{u}'_L
\not\!\!{W^+}d_L+h.c., 
\ee
with
\be
M_d \equiv diag(m_d,m_s,m_b)=U^\dagger\tilde{M}_d V_d
\ee
and
\be
\hat{M}_u\equiv
\left(\bea{ccc} 
0 & m_{12} & m_{13} \\
m_{21} & 0 & m_{23} \\
0 & m_{32} & m_{33}
\ea   \right)
=U^\dagger\tilde{M}_u V_u.
\label{hatm}
\ee
The $m_{ij}$ are complex numbers, $m_{ij}=N\rho_{ij}\exp{(i r_{ij})}$, with
$N=m_t+m_c+m_u$ a suitable normalization constant. 

The two matrices $U$ and $V_d$ are determined by solving the two eigenvalue
problems 
\be
U^\dagger \tilde{M}_d\tilde{M}_d^\dagger U=diag(m^2_d,m^2_s,m^2_b),~~
V_d^\dagger \tilde{M}_d^\dagger\tilde{M}_d V_d=diag(m^2_d,m^2_s,m^2_b), 
\ee
while $V_u$ is chosen in such a way to get the three zeroes in eq.
\eqn{hatm} (we use the following notation: $A_{i.}$ is the i-th row,
$A_{.i}$ is the i-th column of a matrix $A$ and $\times$ represents cross
product): 
\beqa
(V_u)_{.1} &\propto& (U^\dagger \tilde{M}_u)_{1.}\times(U^\dagger
\tilde{M}_u)_{3.} \nn \\ 
(V_u)_{.2} &\propto& (V_u)^*_{.1}\times(U^\dagger \tilde{M}_u)_{2.} \\
(V_u)_{.3} &\propto& (V_u)^*_{.1}\times(V_u)^*_{.2}, \nn
\eeqa
where the multiplicative constants are determined requiring the
$(V_u)_{.i}$ to be norm-one vectors. 

In this way $\hat{M}_u$ contains twelve real parameters, six moduli and six
phases. Depending on the arbitrariness of the differences between quark
phases we have the possibility to remove five of them remaining with one
phase and six moduli, which must be compared with the seven physical
parameters given by the up quark masses and Cabibbo-Kobayashi-Maskawa
angles and phase. 

Obviously, different choices of phases correspond to different
representations of K so, to compare our result with the various
parametrizations of $K$, the phases must be chosen in an appropriate way.
It is possible to reproduce the usual representations of $K$
\cite{padb,wolf}, 
\beqa
K &=& \left(\bea{ccc} 
c_{12}c_{13} & s_{12}c_{13} & s_{13}e^{-i\delta_{13}}  \\
-s_{12}c_{23}-c_{12}s_{23}s_{13}e^{i\delta_{13}}   & 
c_{12}c_{23}-s_{12}s_{23}s_{13}e^{i\delta_{13}}    &
s_{23}c_{13} \\
s_{12}s_{23}-c_{12}c_{23}s_{13}e^{i\delta_{13}}    &
-c_{12}s_{23}-s_{12}c_{23}s_{13}e^{i\delta_{13}}   & 
c_{23}c_{13}
\ea \right),
\label{krep} \\
K &\simeq& \left(\bea{ccc} 
1-\frac{\lambda^2}{2} & \lambda  & \lambda^3A \left[\rho-i\eta(1-
\frac{\lambda^2}{2} )\right] \\ 
-\lambda & 1-\frac{\lambda^2}{2}-i\eta A^2\lambda^4  & \lambda^2A
(1+i\eta\lambda^2) \\ 
\lambda^3A (1-\rho-i\eta) & -\lambda^2 A & 1 
\ea   \right), \nn
\eeqa
by means of the three phases $r_{12}, r_{13}, r_{23}$. Starting from these
ones, by redefining the left and right-handed up quark phases as 
\beqa
(u'_L,c'_L,t'_L) & \longrightarrow & \left( u'_Le^{ir_{13}},
c'_Le^{ir_{23}},t'_L \right) \\ 
(u'_R,c'_R,t'_R) & \longrightarrow & \left(
u'_Re^{ir_{23}},c'_R,t'_R\right), \nn 
\eeqa
we can rotate out all the phases except for the combination
$\Phi=r_{12}-r_{13}$. After these transformations, all the $m_{ij}$ become
real but $m_{12}=N\rho_{12}e^{i\Phi}$. Thus $\Phi$ is the only combination
of phases that has physical relevance. This can be seen calculating, for
example, the imaginary part of the fourth order invariant of $K$
\cite{inva}, 
\be
J\equiv Im\left( \Delta^{(4)}_{\alpha\rho} \right)\equiv Im\left(
K_{\beta\sigma} K_{\gamma\tau} K^{*}_{\beta\tau} K^{*}_{\gamma\sigma}
\right)= s_1^2s_2s_3c_1c_2c_3 \sin{\delta}\simeq\lambda^6A^2\eta 
\ee
(no summation on repeated indices is intended and $\alpha,~\beta,~\gamma,~
(\rho,~\sigma,~\tau)$ cyclic), where the two last terms refer to the
representations of $K$ given in eq. \eqn{krep}. The expression of $J$ found
within our representation is given (after having calculated $K$) in eq.
\eqn{jei}. 

With another change of basis,
\be
\left\{ \bea{ c }  u'_L=S_L u_L \\ u'_R=S_R u_R \ea \right.,
\ee
we obtain
\be
L=\bar{u}_L{M}_{u}u_R+\bar{d}_L{M}_{d}d_R+g\bar{u}_L\not\!\!{W^+}
S^\dagger_Ld_L+h.c., 
\ee
where $S_L$ and $S_R$ are chosen to diagonalize $\hat{M}_u$,
\be
M_u\equiv  diag(m_u,m_c,m_t)= S_L^\dagger\hat{M}_u S_R.
\ee
Consequently $K$ is given by $K_{ij}=(S_L^\dagger)_{ij}$ where $S_L$ is the
unitary matrix which diagonalizes the product $\hat{M}_u\hat{M}_u^\dagger$.
We find for the eigenvectors of $\hat{M}_u\hat{M}_u^\dagger$ the following
expression: 
\be
u_i=\left(\alpha_i,\beta_i,\gamma_i \right)/\sqrt{|\alpha_i|^2+
|\beta_i|^2+|\gamma_i|^2}, 
\ee
with ($l_i=m_i/N,~ i=u,c,t$)
\beqa
\alpha_i &=& (\rho_{21}^2+\rho_{23}^2-l_i^2)
(\rho_{32}^2+\rho_{33}^2-\l_i^2 ) -\rho_{23}^2 \rho_{33}^2
\label{vec} \\
\beta_i  &=& (l^2_i-\rho_{32}^2)
\frac{m_{23}m^*_{13}}{N^2}+\frac{m_{23}m_{32}m^*_{12}m^*_{33}}{N^4} \nn \\ 
\gamma_i  &=&
(l^2_i-\rho_{21}^2)\frac{(m_{32}m^*_{12}+m_{33}m^*_{13})}{N^2} -
\rho_{23}^2\frac{m_{32}m^*_{12}}{N^2}, \nn
\eeqa
so that $K_{ij}=(u^*_i)_j$. $J$ is given by
\be
J=\frac{\alpha^{*}_u \alpha_c(l_u^2-l^2_c)} {
(|\alpha_u|^2+|\beta_u|^2+|\gamma_u|^2)(|\alpha_c|^2+|\beta_c|^2+|\gamma_c|^2)
} \rho_{12}\rho_{13}\rho_{23}^2\rho_{32}\rho_{33}\sin{\Phi}. 
\label{jei}
\ee

In the basis given by eq. \eqn{bas1} it is very easy to obtain the mass
matrix $\hat{M}_u$ as a function of $K$ and of the quark masses; indeed we
have 
\be
\hat{M}_u\hat{M}^\dagger_u=K^\dagger  diag(m^2_u,m^2_c,m^2_t) K \equiv N^2
(a_{ij}+i b_{ij}) 
\ee
($i,j=1,2,3$ and $b_{ii}=0$), that is
\beqa
& & \left(\bea{ccc} 
\rho_{12}^2+\rho_{13}^2 & \rho_{13}\rho_{23}e^{i(r_{13}-r_{23})} &
\rho_{12}\rho_{32}e^{i r_{12} }+\rho_{13}\rho_{33}e^{i r_{13} } \\
\rho_{13}\rho_{23}e^{-i(r_{13}-r_{23})}   & \rho_{21}^2+\rho_{23}^2   &
\rho_{23}\rho_{33}e^{i r_{23}} \\ 
\rho_{12}\rho_{32}e^{-i r_{12} }+\rho_{13}\rho_{33}e^{-i r_{13} } &
\rho_{23}\rho_{33}e^{-i r_{23}} &
\rho_{32}^2+\rho_{33}^2 \ea \right) = \nn \\ 
& = &
\left(\bea{ccc} 
a_{11} & a_{12}+ib_{12} & a_{13}+ib_{13} \\
a_{12}-ib_{12}  & a_{22} & a_{23}+ib_{23} \\
a_{13}-ib_{13}  & a_{23}-ib_{23} & a_{33} 
\ea \right).
\label{solu}
\eeqa
Given a particular representation of $K$ these are the most general
equations relating mass matrix and physical parameters. Depending on the
phase choice one can reduce the three imaginary equations in \eqn{solu} to
just one, but we keep all of them to allow any arbitrary phase convention
in $K$. Solving eq. \eqn{solu} we find: 
\beqa
\tan{r_{12}} = \frac{ b_{13}\rho^2_{23}-a_{12}b_{23}-a_{23}b_{12}}
{a_{13}\rho^2_{23}-a_{12}a_{23}+b_{12}b_{23}}
& & \tan{r_{13}} = \frac{ a_{23}b_{12}+a_{12}b_{23} } {
a_{12}a_{23}-b_{12}b_{23} } 
\label{rho} \\
\tan{r_{23}}  =  \frac{b_{23}}{a_{23} }  & & 
\rho_{12} = \pm\sqrt{ a_{11}-\frac{ a_{12}^2+b_{12}^2 }{\rho_{23}^2} } \nn \\ 
\rho_{13} = -\frac{ \sqrt{a_{12}^2+b_{12}^2 } }{\rho_{23} } & & 
\rho_{21} = \pm\frac{l_ul_cl_t}{ \sqrt{ a_{11}a_{33}  - a_{13}^2 -b^2_{13} 
} } \nn \\ 
\rho_{23} = \pm\sqrt{a_{22}-\frac{l^2_ul^2_cl^2_t}{a_{11}a_{33}-
a_{13}^2-b_{13}^2} } & & 
\rho_{33} = - \frac{ \sqrt{a_{23}^2+b_{23}^2} }{\rho_{23} } \nn \\ 
\Delta=\rho_{12}(a_{13}\rho_{23}^2 -a_{12}a_{23}+b_{12}b_{23} ) & & 
\rho_{32} = sign(\Delta )
\sqrt{a_{33}-\frac{a_{23}^2+b_{23}^2}{\rho_{23}^2} }. \nn 
\eeqa
Similar formulae hold for the lepton Dirac masses if the exchanges
$(d,s,b)\rightarrow(e,\mu,\tau)$ and
$(u,c,t)\rightarrow(\nu_e,\nu_\mu,\nu_\tau)$ are performed.

For the sake of utility we quote here the expressions for the
$\rho_{ij}^2$, obtained when the Wolfenstein parametrization for $K$
\cite{wolf} is used. These expressions are approximated up to the fifth
order in $\lambda$ and neglecting $l_u^2~ (l_c^2)$ with respect to
$l_c^2~(l_t^2)$: 

{
\footnotesize
\beqa
\rho^2_{12}\simeq l_c^2\lambda^2 \left[ (\eta^2+\rho^2)(1-\lambda^2)+
\rho\lambda^2 \right] & & 
\rho^2_{13}\simeq l_t^2A^2\lambda^6 \left[ 1+\eta^2-\rho(2-\rho) \right]
\nn \\ 
\rho^2_{21}\simeq \frac{l_u^2}{\lambda^2} \left[ 1-2 A^2 \lambda^4
(1-\rho+\eta^2\lambda^2) \right] & & 
\rho^2_{23}\simeq l_t^2A^2\lambda^4
\label{rhoapp} \\ 
\rho^2_{32}\simeq \frac{l_c^2 }{A^2\lambda^4}\left\{ 1-\lambda^2+\lambda^4
\left[ \frac{1}{4}+A^2\left( 2-\lambda^2+A^2\lambda^4 (1+\eta^2)
\right)\right] \right\} & & 
\rho^2_{33}\simeq l_t^2 \nn \\ 
\tan{r_{12}} \simeq \frac{\eta}{4\rho^3} \left[
\lambda^4(\rho-1-4A^2\rho^2)+2\rho(\lambda^2-2\rho) \right] & & 
\tan{r_{13}} \simeq \frac{\eta}{1-\rho} \nn \\ 
\tan{r_{23}} \simeq -\frac{l_c^2}{2 l_t^2} \eta\lambda^2 \left[
2-\lambda^2(1-2A^2) \right]. & & \nn
\eeqa
}

{\small
\begin{table}[t]
\begin{center}
TABLE 1 \\
\vspace{.6truecm}
\begin{tabular}{|c|c|c|c|c|c|c|c|} \hline\hline
\rule[-0.4truecm]{0mm}{1truecm} 
& $\rho_{12}^2$   &  $\rho_{13}^2$   & $\rho_{21}^2$   & $\rho_{23}^2$   & 
$\rho_{32}^2$   &  $\rho_{33}^2$ & J \\ \hline \hline 
\rule[-0.4truecm]{0mm}{1truecm}
exact & $9.54\cdot10^{-8}$ & $8.10\cdot10^{-5}$ & $3.07\cdot10^{-9}$ &
$1.68\cdot10^{-3}$  & $7.54\cdot10^{-3}$ & $0.985$  & $2.9\cdot10^{-5}$ \\
\hline 
\rule[-0.4truecm]{0mm}{1truecm}
approx.  & $9.54\cdot 10^{-8}$  & $8.04\cdot 10^{-5}$  & $3.07\cdot
10^{-9}$  & $1.67\cdot 10^{-3}$  & $7.60\cdot 10^{-3}$  & $0.993$ &
$3.3\cdot10^{-5}$ \\
\hline\hline 
\rule[-0.4truecm]{0mm}{1truecm}
& $\sigma_{12}^2$   &  $\sigma_{13}^2$   &  $\sigma_{21}^2$   &
$\sigma_{23}^2$   &  $\sigma_{32}^2$   &  $\sigma_{33}^2$ & J \\
\hline\hline 
\rule[-0.4truecm]{0mm}{1truecm}
exact & $3.96\cdot10^{-6}$ & $1.11\cdot10^{-4}$ & $3.96\cdot10^{-5}$ &
$2.26\cdot10^{-3}$  & $0.283$ & $0.660$  & $2.9\cdot10^{-5}$ \\
\hline\hline 
\end{tabular}
\end{center}
\end{table}
}

In table 1 both the exact and approximate values are reported. They are
obtained using the central values of the measured ranges of $K$ and the
following quark masses evaluated at $m_{Z^o}=91.187~ GeV$:
$m_t=180~GeV,~m_c=0.661~GeV, m_u=0.00222~GeV$ \cite{fuko}. We quote also
the values of the elements of the down quark matrix when
$M_u=diag(m_u,m_c,m_t)$ and $(\hat{M}_d)_{ij}=N'\sigma_{ij}e^{is_{ij}}$
($N'=m_d+m_s+m_b,~m_d=0.00442~GeV,~m_s=0.0847~GeV,~m_b=2.996~GeV$). We did
not report the results for the $\sigma_{ij}$'s when the approximate
formulae \eqn{rhoapp} are used since the weaker hierarchy between down
quark masses makes it necessary a higher order approximation. 

In conclusion, we exhibit a representation for the fermion mass matrices
where the expression of the Cabibbo-Kobayashi-Maskawa matrix is relatively
easy. We solve the problem of inverting the relationship between the mass
matrices and physical parameters. The manageable formulae we find can be
useful in investigating the various hypotheses formulated on this sector of
the {\it Standard Model}. 

We are very much indebted with F. Buccella for his suggestions and  for the
many discussions we had. We also thank G. Mangano for discussions and
Zhi-zhong Xing for useful comments. We are very grateful to P. Vitale for
her support.

\end{document}